\documentclass[12pt,onecolumn,preprint]{revtex4}
\usepackage{amssymb}
\usepackage{amsmath}
\usepackage{graphicx}
\usepackage{mathrsfs}
\usepackage{color}
\usepackage{ulem}
\usepackage{setspace}

\begin{document}

\title{Dynamics of vortex quadrupoles in nonrotating trapped Bose-Einstein condensates}
\author{Tao Yang$^{1,2,*}$}
\author{Zhi-Qiang Hu$^1$}
\author{Shan Zou$^3$}
\author{Wu-Ming Liu$^{4,\dagger}$}

\affiliation{$^1$Institute of Modern Physics, Northwest University, Xi'an, 710069,  China\\
$^2$Shaanxi Key Laboratory for Theoretical Physics Frontiers, Xi¡¯an, 710069, China\\
$^3$School of Physics, Northwest University, Xi'an, 710069, China\\
$^4$Beijing National Laboratory for Condensed Matter Physics, Institute of Physics, Chinese Academy of Sciences,
Beijing 100190, China}
\email{ yangt@nwu.edu.cn, wliu@iphy.ac.cn}

\begin{abstract}
  Dynamics of vortex clusters is essential for understanding diverse superfluid phenomena. In this paper, we examine the dynamics of vortex quadrupoles in a trapped two-dimensional (2D) Bose-Einstein condensate. We find that the movement of these vortex-clusters fall into three distinct regimes which are fully described by the radial positions of the vortices in a 2D isotropic harmonic trap, or by the major radius (minor radius) of the elliptical equipotential lines decided by the vortex positions in a 2D anisotropic harmonic trap. In the ``recombination'' and ``exchange'' regimes the quadrupole structure maintains, while the vortices annihilate each other permanently in the ``annihilation'' regime. We find that the mechanism of the charge flipping in the ``exchange'' regime and the disappearance of the quadrupole structure in the ``annihilation'' regime are both through an intermediate state where two vortex dipoles connected through a soliton ring. We give the parameter ranges for these three regimes in coordinate space for a specific initial configuration and phase diagram of the vortex positions with respect to the Thomas-Fermi radius of the condensate. We show that the results are also applicable to systems with quantum fluctuations for the short-time evolution.
\end{abstract}

\maketitle

\newpage

\section*{Introduction}

Vortices could be observed in most realm of physics such as hydrodynamics, superfluids, optical fields and even cosmology. The dynamics of quantized vortices is essential for understanding diverse superfluid phenomena such as critical-current densities in superconductors\cite{PRB.68.224501}, quantum turbulence\cite{PNAS.111.4647,PR.524.85,JLTP.128.167,PoF.1.3678335,PRL.91.194502} and novel quantum phases\cite{RMP.66.1125,PR.355.235,PRL.87.120405,PRL.100.130402,RMP.85.299} in superfluids. Vortices are also topological defects that play key roles in transport, dissipative and coherent properties \cite{PRA.84.011605,PRA.68.063609,PRB.72.014546,JPB.43.155303,PRA.89.043613} of superfluid systems. The pioneering work of Yarmchuk \textit{et al.} in 1979 successfully located the ends of parallel vortex lines in superfluid Helium \cite{PRL.43.214}. The real-time dynamics of vortex lattice in type II super conductors was observed in 1992 \cite{Nature.360.51}. Until 2006, direct observation of quantized vortex lines in superfluid Helium in arbitrary three-dimensional configurations has been achieved \cite{Nature.441.588}.

The realization of Bose-Einstein condensates (BECs) provides an accessible and highly controllable platform for fundamental studies of superfluid vortex dynamics, and has been followed by various theoretical investigations and numerical analyses. It is remarkable that, comparing with other systems ruled by nonlinear Schr\"{o}dinger equations, BECs are the ideal laboratory for finding these nonlinear excitations due to larger interaction strengths and easier tunable parameters. The size of vortex cores in a BEC is proportional to the healing length of the condensate, $\zeta=\hbar/\sqrt{2mg\rho}$, where $m$ is the mass of atoms, $\rho$ is the density of the condensate in the absence of vortex, and $g$ is the interatomic interaction strength. For given trap frequencies, $g$ is proportional to the scattering length between atoms, which can be easily adjusted by using magnetic or optical Feshbach resonances \cite{PRL.81.69,Nature.392.151,PRL.93.123001}. Due to the matter wave nature of condensates \cite{Science.275.637}, vortices can be detected in atomic interference. Matthews \textit{et al.} \cite{PRL.83.2498} firstly demonstrated vortex production through an interference measurement, and later, Inouye \textit{et al.} observed vortex phase singularities as dislocations in the interference fringes in BECs \cite{PRL.87.080402}. As the size of the vortex cores in a trapped condensate is ordinarily several times smaller than the wavelength of light used for imaging, in experiments the condensates are usually allowed to expand to a point at which the vortex cores are large compared to the imaging resolution\cite{PRL.83.2498,PRL.84.806,PRL.87.080402,PRA.70.063607,PRA.82.033616,PRL.94.040403,PRL.103.045301}. Many works have been done to develop the vortex detection techniques and to understand vortex dynamics during the interference of BECs \cite{PRL.81.5477,PRL.87.080402,PRA.64.031601,SSC.108.993,PRA.87.023603,PRA.88.043602,LP.24.115502}, which is important for the applications of matter-wave interferometry. However, the direct, \textit{in situ} observation of vortices in a trapped condensate without expansion was achieved experimentally only in the past few decades. With the development of atom chip technology \cite{RMP.79.235} and recent experimental achievements in the monitoring of the dynamics of vorticity states in BEC systems in real time \cite{Science.329.1182}, there has been fast-growing interest in vortex dynamics, and the new physical effects associated with the trap geometry.

Condensates can afford a rich variety of vortex-cluster structures, including soliton-vortex hybrids \cite{PRL.94.040403}, vortex chains \cite{PRA.87.063630,PRB.66.014509},turbulent vortex tangles \cite{PRA.86.013631}, corotating vortices\cite{PRL.110.225301} vortex dipoles \cite{PRL.104.160401,PRA.68.063609,OL.26.1601,PRA.77.053610,CPAA.10.1589,SIAM.14.699,PLA.375.3044}, vortex tripole \cite{PRA.71.033626,PRA.74.023603} and vortex quadrupoles \cite{PRA.61.032110,PRA.65.023603,PRE.66.036612,PRA.82.013646,Phys.D.240.1449}. Different from the vortex lattice induced by the rotation of BECs\cite{PRA.84.053607,PRA.85.043626,SR.4.4224,SR.6.19380}, the complex structures made by vortices with opposite charges are excited collective states of nonrotating BECs. They are pure nonlinear entities presenting in a system, and can maintain stable topological states. When the interactions between them, rather than variations in density, dominate the dynamics, the dynamics of the system is rather complicated. The stability properties of a single vortex system and some vortex structures in nonrotating BECs were discussed in Refs. \cite{PRL.84.5919,PRA.68.063609,PRA.74.023603,PRE.66.036612,PRA.61.032110,PRA.70.043619,PRA.71.033626,
PRA.81.061805} with respect to interatomic interaction strength and trap anisotropy.

A vortex quadrupole is a vortex cluster with zero total vorticity constructed as the product of four singly charged vortices.
Previous study showed that this cluster configuration is both energetically and dynamically
unstable \cite{PRA.71.033626}. However, for some special configurations, stationary vortex quadrupoles were found to exist \cite{PRE.66.036612,PRA.71.033626}. It has shown that a vortex quadrupole is always structurally unstable against perturbations in some external trap parameters \cite{PRL.87.140403}. But a complete picture of quadrupole dynamics has not been given, which provides critical information for understanding how the vortices influence superfluid behaviors, such as evolution to turbulence and decay processes.

In this paper we study, by numerical simulations, the detailed dynamics of two-dimensional (2D) vortex quadrupoles with different initial configurations in trapped BECs, and for the first time we identify the parameter ranges which give the distinct regimes for the revival and disappearance of a quadrupole structure. The paper is organized as follows. In theoretical model section we introduce the model under investigation and define the general initial state of a vortex quadrupole. The method to introduce quantum fluctuations into the systems is also presented. In numerical results and analysis section the trajectory of each vortex in a quadrupole is investigated in detail for isotropic, anisotropic and fluctuating condensates. The mechanism of dipole formation and annihilation is discussed. The ranges of parameters of the initial configurations of quadrupoles corresponding to different dynamical regimes are calculated. A brief conclusion is also given.


\section*{Theoretical model}\label{sec1}
In the frame of zero temperature mean-field theory, a trapped BEC system is well parameterized by $\psi(x,y,z,t)$ which satisfies the Gross-Pitaevskii (GP) equation
\begin{equation}\label{eq1}
  i\hbar\partial_{t}\psi=\left[-\frac{\hbar^2}{2m}\nabla^2+\frac{1}{2}m(\omega_{x}^{2}x^{2}+\omega_{y}^{2}y^{2}+\omega_{z}^{2}z^{2})+g_{3D}N|\psi|^{2}\right]\psi,
\end{equation}
with $\omega_j$ being the trap frequency and $g_{3D}={4\pi\hbar^2a_s}/{m}$ being the three-dimensional coupling constant, where $\psi$ is normalized to 1. For a large aspect ratio of the trap potential $\omega_z\gg\omega_x=\omega_y=\omega_\perp$, the condensate can be regarded as a 2D disk-shaped cloud, where the motion in the $z$-direction is squeezed. The 2D order parameter $\psi(x,y,t)$ is governed by the corresponding 2D-GP equation, where the coupling constant $g_{3D}$ in Eq.(\ref{eq1}) is substituted by $g_{2D}=g_{3D}/\sqrt{2\pi}a_z$ with $a_z=\sqrt{\hbar/m\omega_z}$. In our simulations of a $^{87}Rb$ BEC containing $N=1.5\times10^5$ atoms, the bulk $s$-wave scattering length is $a_s=5.4$\textrm{nm}. For the trap frequencies, we used a 2D disk-shaped configuration, where $\omega_x=\omega_y=2\pi\times5$ \textrm{rad}$\cdot$\textrm{s}$^{-1}$ and $\omega_z=2\pi\times100$ \textrm{rad}$\cdot$\textrm{s}$^{-1}$. It is convenient to turn all quantities dimensionless by introducing a time scale $t_0={1}/{\omega_{\perp}}$ and a length scale $a_{0}=\sqrt{{\hbar}/{m\omega_{\perp}}}$, which can provide better precision in simulations. For these parameters, the 2D Thomas-Fermi (TF) radius of the condensate is $R_{TF}=\left(8\sqrt{2}a_sN_0/\sqrt{\pi}a_z\right)^{1/4}\approx8.32a_0$.

The order parameter of the system can be well described by a single particle wave function $\psi(\textbf{r},t)=\sqrt{\rho(\textbf{r},t)}\exp[i\theta(\textbf{r},t)]$, where $\rho(\textbf{r},t)$ and $\theta(\textbf{r},t)$ are density and phase distributions of atoms respectively, with $\textbf{r}=(x,y)$. Vortices can be seeded into the system by employing the phase imprinting techniques \cite{PRL.83.2498,Nature.401.568} experimentally. Numerically, the quadrupole states of interest in our simulations are obtained by using the method in Refs.\cite{PRA.79.053615} and \cite{PRA.86.013631}. Firstly the vortex-free ground state of the condensate, $\psi_g$, is obtained by the imaginary time evolution. The initial phase factor of the order parameter of the quadrupole state is
\begin{equation}
  \theta(\textbf{r},t=0)=\prod_js_j\arctan\left(\frac{y-y_{j}}{x-x_{j}}\right),
\end{equation}
where $s_j=\pm 1$ is the topological charge of the $j_{th}$ vortex, imprinting the velocity field of a vortex located at $(x_{j},y_{j})$. The radial position of each vortex is then $r_j=\sqrt{x_j^2+y_j^2}$. With new initial condition $\psi_g\exp{[i\theta(\textbf{r},t=0)]}$, the quadrupole vorticity state of a BEC system can be  obtained by evolving the GP equation again in imaginary time while forcing the phase distribution during the calculation to be $\theta(\textbf{r},t=0)$ until the solution is numerically well converged, i.e. the relative change in norm of number of atoms and in energy of the system is smaller than $10^{-6}$. The converged quasi-equilibrium state is then the desired quadrupole configuration ready for a real-time evolution. In this paper we focus on a subset of possible configurations, namely, the initial vortex quadrupole structures are axisymmetric. An example density profile and the corresponding phase diagram are shown in Fig. \ref{fig1}e, where the vortices with opposite charges are nested in the condensate alternately forming dipole structures along each axis.  

The thermal and quantum fluctuations are experimentally only important for temperatures which are larger or comparable to the mean-field interactions. Since we are studying systems in the frame of zero temperature mean-field theory, it is reasonable to think that these fluctuations will not affect the dynamics significantly, especially during the short-time evolution. However, taking the symmetry of systems into account, quantum fluctuations might have obvious effects on the long-time evolution of the systems. To introduce quantum fluctuations into our systems, the coordinate space order parameter equivalent to our choice of stochastic initial state is easily shown to be
\begin{equation}
  \Psi({\bf{r}}, t=0)=\psi_0({\bf{r}})+\xi({\bf{r}})
\end{equation}
Where $\psi_0({\bf{r}})$ is chosen to be the initial state containing the quadrupole structure used in the situations without fluctuations, i.e. wave function of the condensed atoms.  $\xi({\bf{r}})$ is a complex Gaussian random field, which represents the uncorrelated quantum fluctuations\cite{PRL.94.040401,JPB.46.145307}
\begin{align}
 \langle\xi({\bf{r}})\rangle&=\langle\xi^*({\bf{r}})\rangle=0,\\
 \langle\xi({\bf{r}})\xi({\bf{r}}')\rangle&=\langle\xi^*(x)\xi^*({\bf{r}}')\rangle=0,\\
 \langle\xi^*({\bf{r}})\xi({\bf{r}}')\rangle&=\frac12\delta({\bf{r}}-{\bf{r}}').
\end{align}
In practice, one can implement this initial condition by utilizing a complete set of plane-wave basis functions $\{\phi_j({\bf{r}})\}=\frac{1}{\sqrt{V}}e^{i{\bf{k}}_j\cdot{\bf{r}}}$,
such that the random field can be written as
  \begin{equation}
   \xi({\bf{r}})=\sum_j\xi_j\phi_j({\bf{r}}).
  \end{equation}
Here $\xi_j$ are complex gaussian random variables 
\begin{align}
 \langle\xi_j\rangle&=\langle\xi^*_j\rangle=0,\\
 \langle\xi_j\xi_{j'}\rangle&=\langle\xi^*_j\xi^*_{j'}\rangle=0,\\
 \langle\xi^*_j\xi_{j'}\rangle&=\frac12\delta_{jj'}.
\end{align}
By using the truncated Wigner method \cite{PRA.58.4824,JPB.35.3599}, and integrating the fully coordinate space stochastic differential equation of a 2D system
\begin{equation}\label{TWA}
  i\hbar\frac{\partial\Psi({\bf{r}},t)}{\partial t}=-\frac{\hbar^2}{2m}\nabla^2\Psi({\bf{r}},t)
  +\big[V_{trap}({\bf{r}})+g_{2D}|\Psi|^{2}\big]\Psi({\bf{r}}, t)
\end{equation}
in real-time, the dynamics of systems under the influence of quantum fluctuations can be explored. The normalisation condition is defined as $N_T=\int d\textbf{r}|\Psi(\textbf{r}, t)|^{2}$ which includes both the condensed atoms and the virtual particles introduced into the field. It is obvious that even though Eq.(\ref{TWA}) has a form similar to that of the GP equation (\ref{eq1}), the important difference is that quantum fluctuations (virtual particles) are included in this stochastic differential equation while the GP equation is only valid for the condensed atoms.

In our calculations, only the `low-energy' modes in momentum space contribute to the fluctuations, which means the noise is just added into the modes with amplitude $\xi_{j}, j\in L$, where $L=[-25, 25]$ is a subspace containing $51\times51$ modes, only about $1\%$ of the $501\times501$ momentum space used for our simulations. All other $\xi_j, j\notin L$ are set to be 0.

\section*{Numerical results and analysis}\label{sec2}

To get insight into the dynamics of vortex quadrupoles in a disk-shaped BEC, we first review the motion of a single vortex and a vortex dipole in such a system. A 2D disk-shaped condensate in an axisymmetric harmonic trap is a nonuniform system. In the TF limit, for a single vortex generated in this nonrotating system the precession velocity is a function of the vortex displacement $r$ from the centre of the condensate as $v=\omega_mr/(1-r^2/R_\perp^2)$, where $\omega_m=(3\hbar/2 mR_\perp^2)\ln(R_\perp/R_c)$ is the metastable frequency for the appearance of a central vortex in a disk-shaped condensate \cite{JPCM.13.R135}, with $R_{\perp}$ being the mean condensate radius and $R_c$ being the vortex core radius. It is obvious that a central vortex does not undergo any evolution over time. At zero temperature, an off-centre vortex with positive (negative) charge will perform a circular motion around the centre of the condensate in a anti-clockwise (clockwise) direction in a constant radius with a constant energy \cite{PRL.84.5919,PRA.61.063612} without considering the interaction between the vortex and the excited phonon mode. This pattern of behaviour repeats as the density wave oscillations induced by the vortex motion are all confined within the condensate, and no net phonon radiation is expected. However, due to the interaction of the vortex with the collective phonon mode, the vortex trajectory is not a perfect circle \cite{JPCM.13.R135,PRL.92.160403}. At finite temperatures it spiral out of the condensate due to its interaction with the thermal component \cite{PRA.79.053615}.

%

For a dipole structure of vortex-antivortex pair in a trap, the dynamics is the result of the competition between the density inhomogeneity driven and the dipole interaction driven (mutually driven of each vortex of the pair). In the case where the mutually driven and inhomogeneity driven balance with each other, a stationary dipole structure generates. For configurations, where a dipole placed symmetrically about the trap centre, vortices make close orbits about fixed points \cite{CPAA.10.1589}. Nearly the same as the motion in a single vortex system, each of the vortices begins its orbit along the equipotential line as if they move independently, but once the vortices begin to approach each other they begin to interact. Instead of annihilating each other, both vortices alter their paths and begin to travel down the condensate from one side to the other side, parallel to each other. This continues until the vortices begin to reach the boundary of the condensate, then both of them move back on to their original orbits.




The situation gets more complicated when a quadrupole structure is taken into account. By setting different initial configurations, i.e. varying the radial position of the vortices, we find that the dynamics of the vortices shows three distinct regimes, which are named as ``recombination'', ``annihilation'' and ``exchange''.

\begin{figure}
\centering
\includegraphics[scale=0.8]{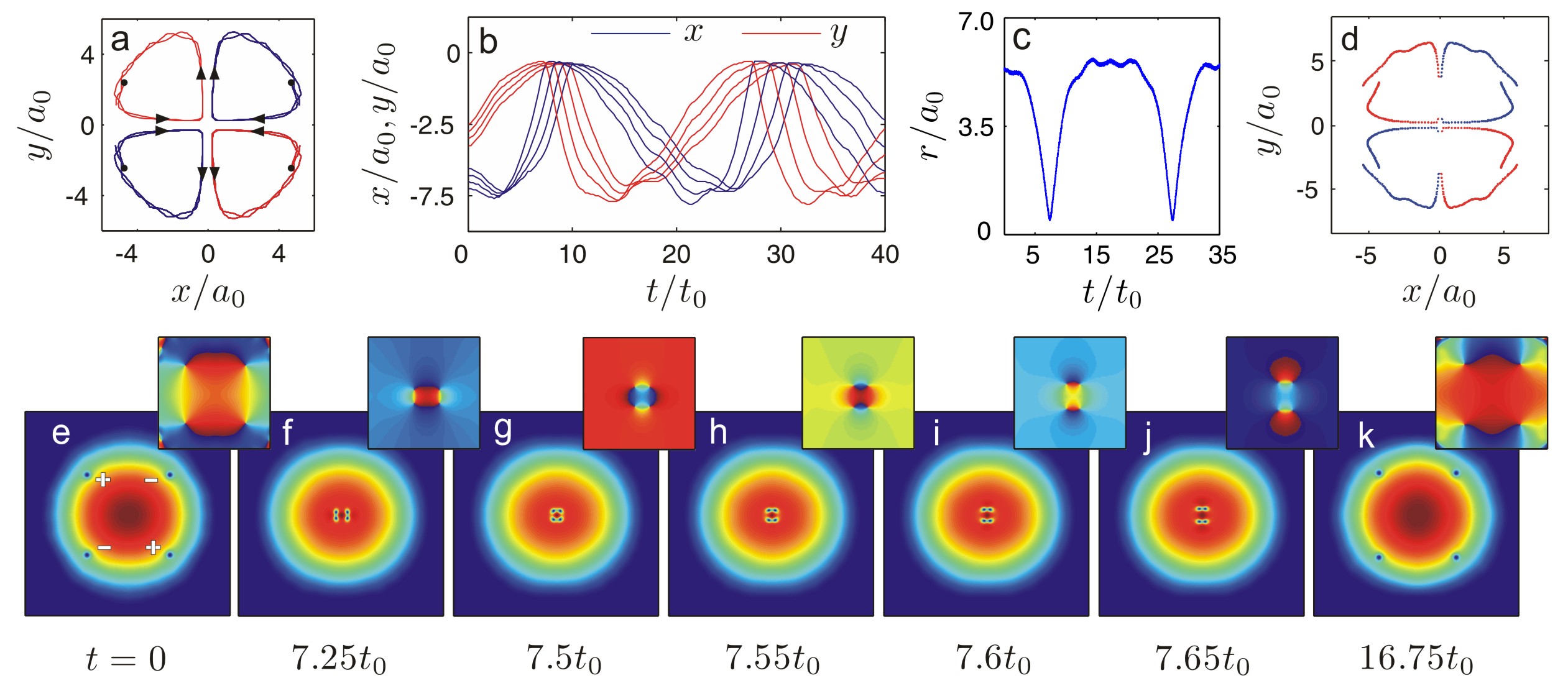} 
\caption{\textbf{ $\arrowvert$ Dynamics of quadrupoles in the ``recombination'' regime.}(a) The trajectory of the quadrupole (black dots) initially seeded at ($\pm 4.75, \pm2.5$) in the condensate. Arrows show the directions of their movements. (b) The time series of the $x$ (blue) and $y$ (red) coordinates of the down-left vortex for four initial configurations, (-4, -3.25), (-4.25, -3), (-4.5, -2.75) and (-4.75, -2.5) respectively. (c) The time series of the radial positions of the initially seeded vortices with charge $s=-1$ (blue dots). The initial configuration is the same as that used in (a). (d) The trajectory of the quadrupole initially seeded at ($\pm 4.75, \pm3.35$) in the condensate. (e)-(k) Typical snapshots for the density distribution and corresponding phase diagram of the condensate. The vortices are initial placed at ($\pm 4, \pm4$). }\label{fig1}
\end{figure}

In the recombination regime as shown in Fig. \ref{fig1}a, the vortices take quarter-circular motion and near-linear motion repeatedly, where the solid dots are the initial positions of the vortices, and the arrows show the direction of movement of each vortex. We can see clearly that, when they approach each other, two vortices in the left half-plane move as a dipole towards the trap centre along the x-axis, while the other two in the right half-plane make a dipole movement as well facing to the left pair. However, after the four vortices reach their minimal separation, two of them in the upper half-plane combine into a new dipole structure and move away the trap centre towards the edge of the condensate. So do the remains in the lower-half plane. The trajectories of the upper and lower dipoles are mirror symmetric to the x-axis. 
During the movement of vortices, the density wave oscillations arise across the condensate, and are severe around the edge of the condensate where the density is low. Therefore, the oscillation of the quarter-circular parts of the vortex trajectories in the low density region is obvious as seen in Fig. \ref{fig1}a-d. For different initial positions the time evolution of the $x-$ and $y-$coordinates of the down-left vortex is plotted in Fig. \ref{fig1}b. Typical density and phase distributions in this regime are plotted in Fig. \ref{fig1}e-k in time series. We note that for given parameters in Fig. \ref{fig1}a-c and e-k four vortices never contact. After the first period the vortices will move away from their initial position a little bit due to the interactions with the density wave oscillations arising in the system, unlike a single vortex or a vortex dipole movement which generates close orbits. In Fig. \ref{fig1}c we plot time series for the radial positions of the $s=-1$ vortices (blue dots). If we increase the initial radial displacement of vortices, the dynamics of the quadrupole can still stay within the recombination regime for the first period. But the dipole annihilation and revival can occur, which makes the deviation of the vortex positions from the initial positions sizable as seen in Fig. \ref{fig1}d ($(x, y)=(\pm4.75 , \pm3.35 )$). During the following time evolution, this may induce a transition of the dynamical regime from recombination to exchange (annihilation).

\begin{figure}
\centering
\includegraphics[scale=0.8]{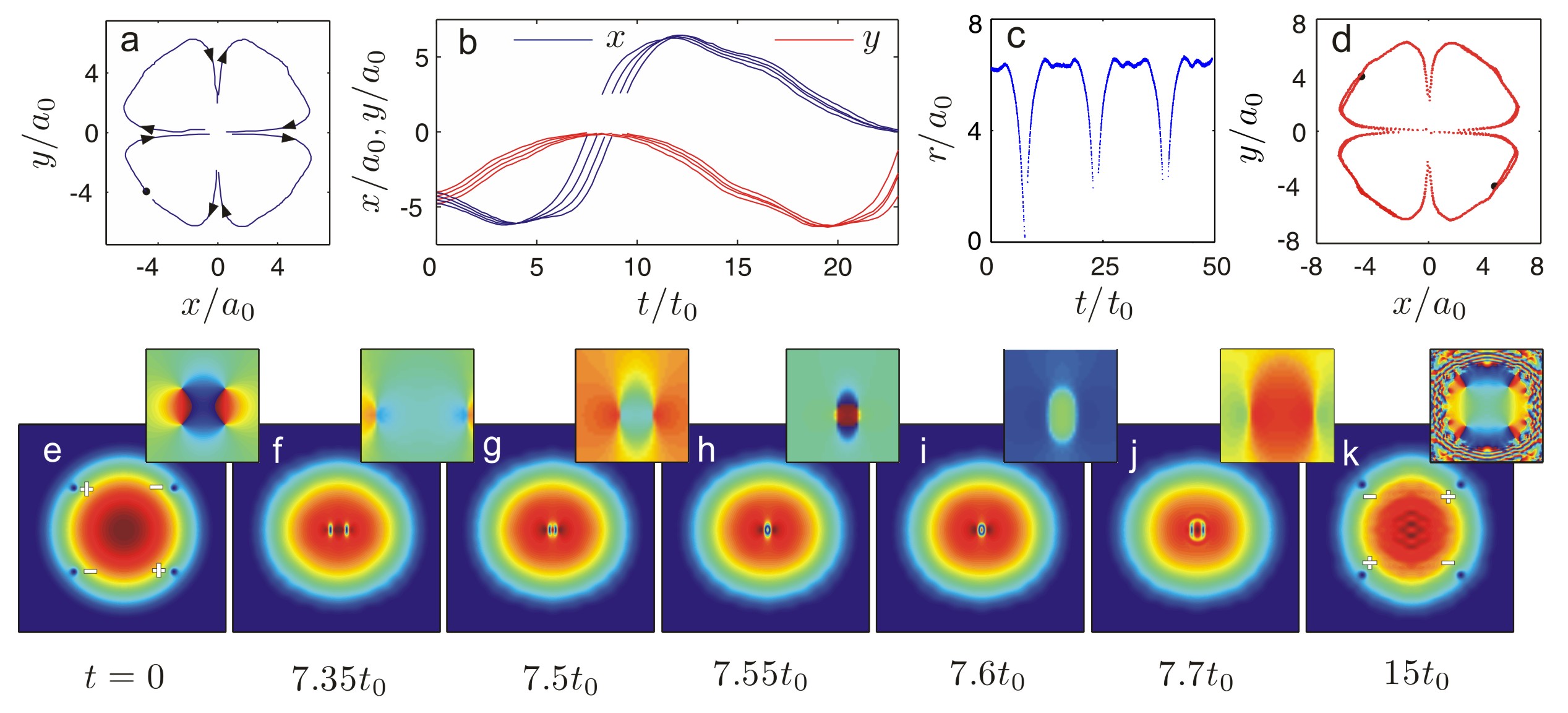}
\caption{\textbf{ $\arrowvert$ Dynamics of quadrupoles in the ``exchange'' regime.}(a) The trajectory of the vortex initially seeded at ($-4.75, -4$) in the condensate. Arrows show the directions of its movement. The Initial positions of four vortices are denoted by the black dots for reference. (b) The time series of the $x$ (blue) and $y$ (red) coordinates of the down-left vortex for four initial configurations, (-4, -4.75), (-4.25, -4.5), (-4.5, -4.25) and (-4.75, -4) respectively. (c) The time series of the radial positions of the initially seeded vortices with charge $s=-1$ (blue dots). (d) The trajectories of the initially seeded vortices with charge $s=1$ (red dots) in the condensate. (e)-(k) Typical snapshots for the density distribution and corresponding phase diagram of the condensate. The initial configurations of (c)-(k) are the same as that used in (a).}\label{fig2}
\end{figure}


\begin{figure}
\centering
\includegraphics[scale=0.8]{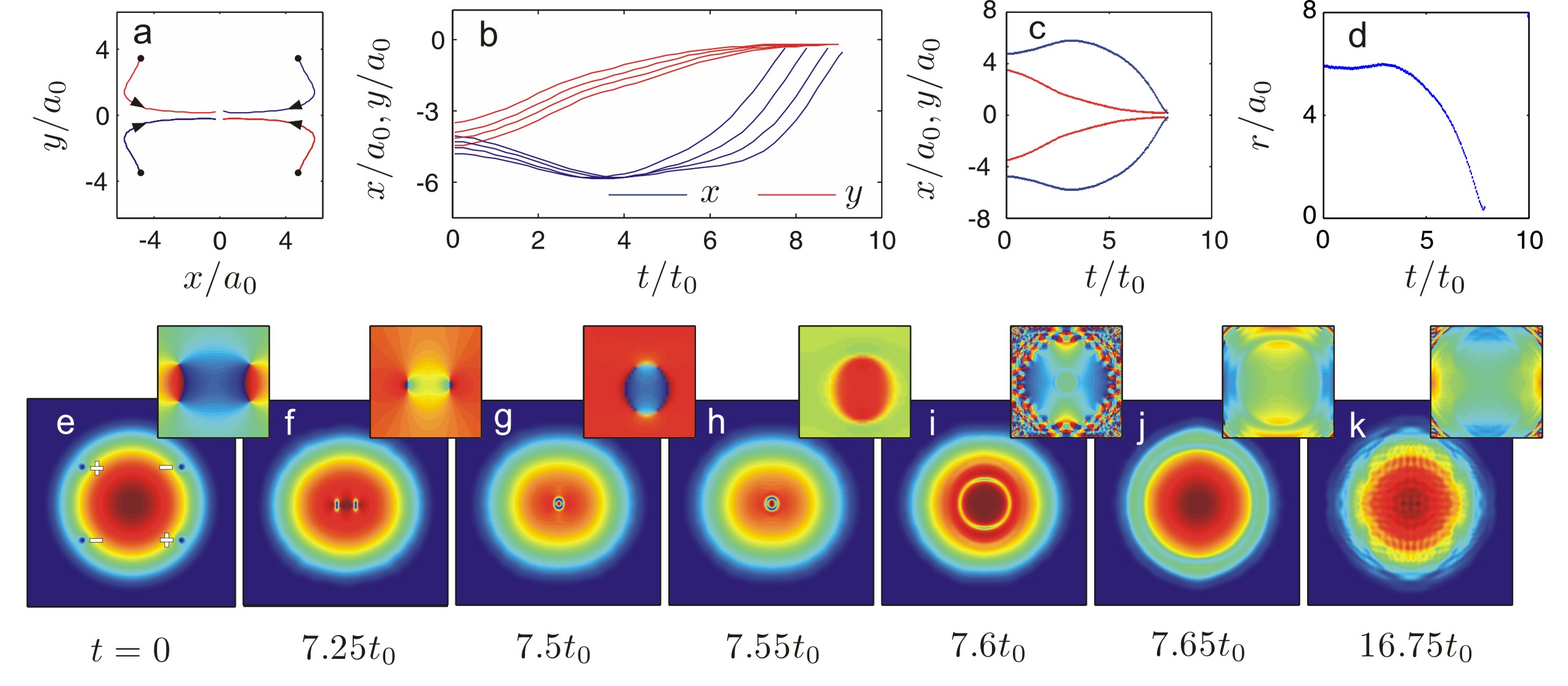}
\caption{\textbf{ $\arrowvert$ Dynamics of quadrupoles in the ``annihilation'' regime.} (a) The trajectories of the vortices initially seeded at ($\pm4.75, \pm3.5$) (black dot) in the condensate. Arrows show the directions of their movements. (b) The time series of the $x$ (blue) and $y$ (red) coordinates of the down-left vortex for four initial configurations, (-4, -4.45), (-4.25, -4.15), (-4.5, -3.9) and (-4.75, -3.5) respectively. (c) The time series of the coordinates of the depicted vortex positions with charge $s=-1$ (initially seeded). Blue and red dots are $x$ and $y$ coordinates vs time respectively. (d) The trajectories of the vortices with charge $s=+1$ initially seeded at ($4.75, -3.5$) and ($-4.75, 3.5$) in the condensate. (e)-(k) Typical snapshots for the density distribution and corresponding phase diagram of the condensate. The initial configuration is the same as that used in (a).}\label{fig3}
\end{figure}
Figure \ref{fig2} shows the trajectory of the down-left vortex (s=-1) for the first period in the exchange regime. From the symmetry of the system we know that the $s=1$ vortices will give the same trajectories but with different starting position and in opposite direction (anticlockwise). Therefore the left dipole and the right dipole move towards each other along the $x$-axis to the trap centre at the beginning, which is the same as the situation in the recombination regime. As they approach each other the distance between the two vortices of the left (right) dipole decreases gradually in the $y-$direction as shown in Fig. \ref{fig2}f. However, instead of separate in the $y$-direction after getting their minimal separation in the recombination regime, the two dipoles join together, and finally form a soliton ring as shown in Fig. \ref{fig2}g-i. This can also be distinguished from the fact that, for some certain time intervals, no vortex trajectories are plotted in Fig. \ref{fig2}b-d. The stability and dynamics of dark soliton rings in condensates has been analyzed in Refs.\cite{PRL.90.120403} and \cite{PRA.77.023625}. However, the dark soliton ring in our system is formed by the connection of four moving vortices during the dynamical process. The soliton ring then breaks into two dipole structures as the original left and right dipoles exchange their positions and then keep moving in the $x$-direction towards the edge of the condensate. After that the vortices in the dipoles separate in the $y$-direction and taking the place of their counterpart. The charge flipping occurs as seen in Fig. \ref{fig2}k as if two dipoles go through each other. The vortices continue to move in a quarter-circular orbit until they approach each other again. But at this time two vortices in the upper half-plane and the other two in the lower half-plane combine into two dipoles respectively. They move in the $y$-direction and repeat what happened in the $x$-direction. After four times charge flipping, the system will restore to its original configuration.


\begin{figure}
\centering
\includegraphics[scale=0.7]{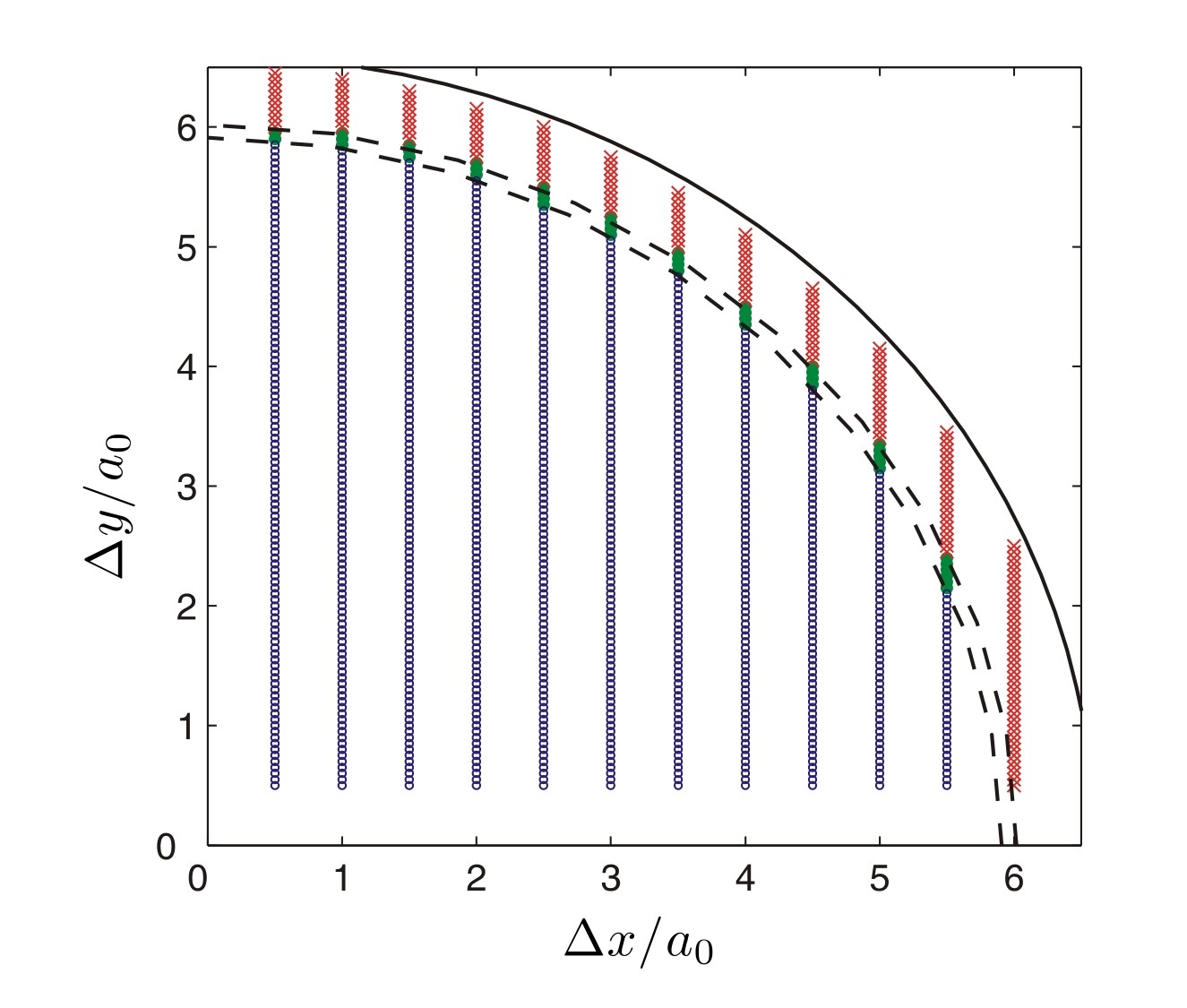}
\caption{\textbf{ $\arrowvert$ Distinguish three dynamical regimes in coordinate space.} Three regions in coordinate space where the dynamics of vortex quadrupoles show whether the vortices will recombine into dipoles (circles), annihilate each other (dots), or exchange their positions (crosses). $\Delta x$, and $\Delta y$ are the initial displacement of the vortices in $x$- and $y$-directions respectively. The dashed-lines show the boundaries between these regions. The solid line is the arc with radius $6.6a_0$. The number of atoms is $1.5\times 10^5$.}\label{fig4}
\end{figure}
The dynamics in the annihilation regime is the most simplest, where the vortices vanish permanently as long as they collide in the trap centre for the first time as shown in Fig. \ref{fig3}a. In this process there is also a ring structure formed. However, different from the situation in the exchange regime, the ring expands across the condensate quickly and smear at the edge of the condensate, which induces severe density wave oscillations as shown in Fig. \ref{fig3}k.

\begin{figure}
\centering
\includegraphics[scale=0.7]{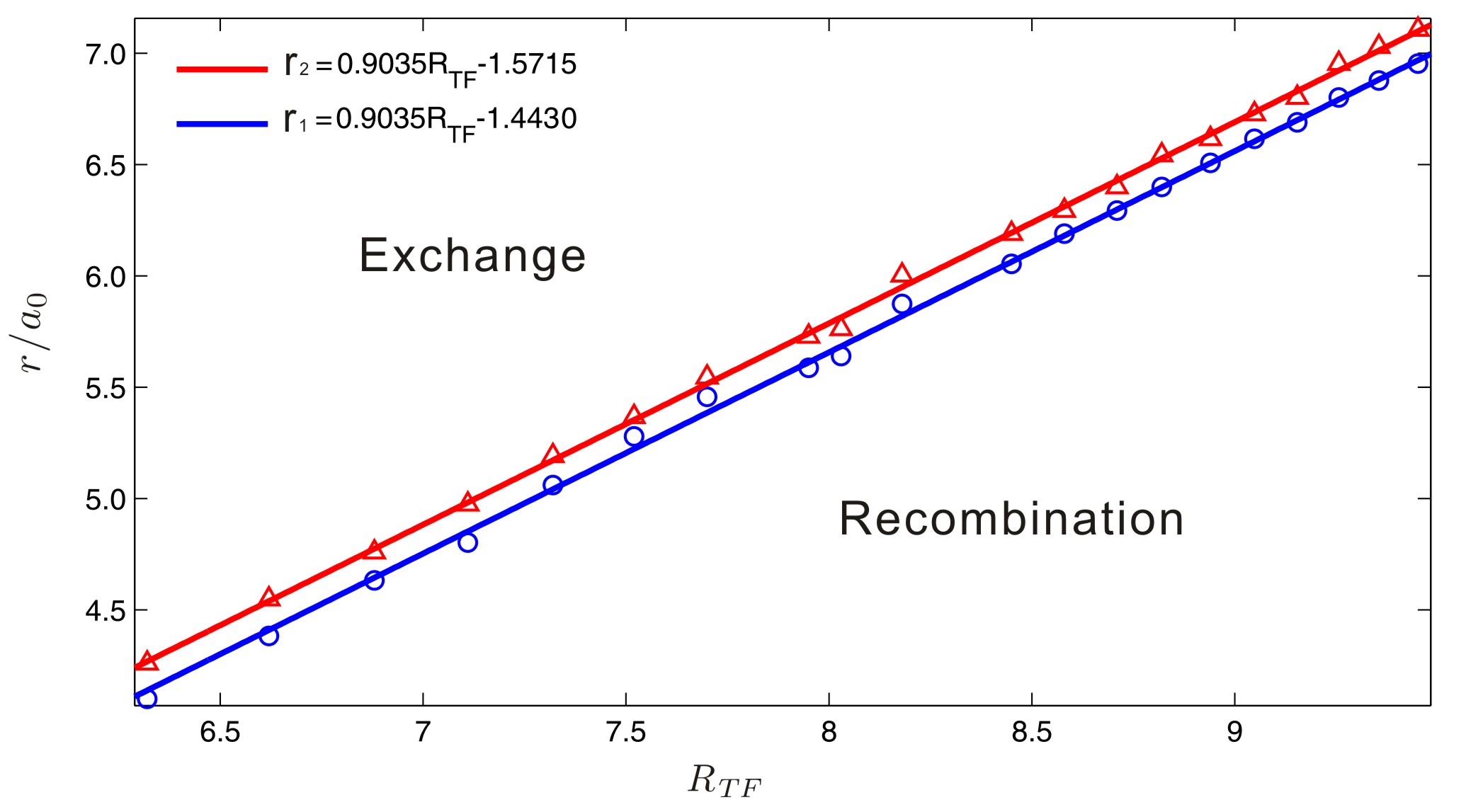}
\caption{\textbf{ $\arrowvert$ Distinguish three dynamical regimes with radial position of vortices.} The boundaries of the three dynamical regimes of vortex quadrupoles as linear functions of the TF-radius. $r$ is the radial position of the vortices in isotropic systems, and the major (minor) radius of the elliptical equipotential lines decided by the vortex positions in anisotropic systems. The circles and triangles are data obtained from the numerical simulations, while the red and blue lines are plotted by the equations shown in the figure legend. The area between the blue and red lines is the annihilation regime.}\label{fig5}
\end{figure}

As is well known, the energy and angular momentum associated with a vortex increase as the vortex is moved to higher density region \cite{JPCM.13.R135}, which coincides with the fact that the energy cost to create a vortex is high when the local density is high \cite{PRA.86.013631}. So we note that in all the three regimes, there are two timescales in the motion: the slow quarter-circular precession around the trap centre and the fast motion nearly parallel to the axis. We note that the density wave oscillations are greater after the annihilation of vortex dipoles and decay of the soliton ring, which indicates that the decay of vortex dipoles or soliton rings can make much severe excitations in dynamical systems. The revival of vortex dipoles can depress the excitations in the system as stated in Refs.\cite{PRA.87.023603,PRA.88.043602}.

In Fig. \ref{fig4} we identify the coordinate parameters which separate regions of ``recombination'' (circles), ``annihilation'' (dots) and ``exchange'' (crosses) dynamics. The boundaries between these regions are both nearly quarter-circles which are denoted by the dashed-lines. For a relatively small radial position $r_{j}<5.91a_0$ the dynamics of the quadrupole is characterized by the combination of the vortices into dipoles in $x-$ and $y-$directions alternately, while for a large radial position $r_{j}>6.02a_0$ exchanging of positions of the vortices and charge flipping occur during the dynamical process. In between these two regions both the quadropole and dipole structures vanish due to the annihilation of the vortices with opposite charges. For $r_j>6.6a_0$ (black solid curve in Fig. \ref{fig4} the density of the condensate is too low to identify a complete vortex structure.

\begin{figure}
\centering
\includegraphics[scale=0.7,bb=700 0 109 300]{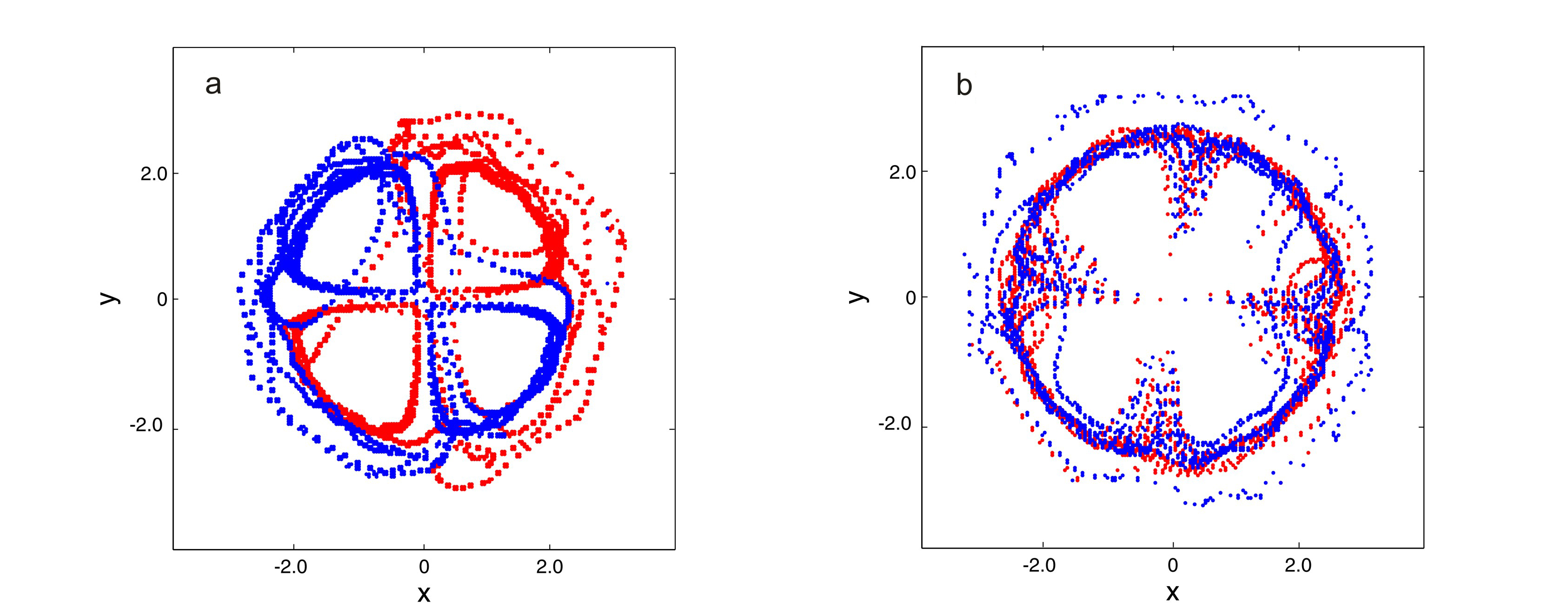}%
\caption{\textbf{ $\arrowvert$ Dynamics of quadrupoles in the systems with quantum fluctuations.} The blue and red dots show the trajectories of vortices with charge $s=1$ and $s=-1$, respectively. (a) The trajectories of the vortices initially seeded at ($\pm4.75, \pm2.5$) in the condensate. The total evolution time is $125t_0$. The chaotic motion of vortices starts from about $86t_0$. (b)The trajectories of the vortices initially seeded at ($\pm4.75, \pm4.0$) in the condensate. The total evolution time is $167t_0$. The chaotic motion of vortices starts from about $128t_0$.}\label{fig6}
\end{figure}
In Fig. \ref{fig5}, we give the radial position indicating the boundaries of three dynamical regimes with respect to varying TF-radius, which is applicable at any time step during the dynamical process. We find that the fitting curves of the boundaries (the blue and red lines) are both proportional to $R_{TF}$.

Through the analysis of isotropic systems, we know that the quarter-circular precessions are nearly along the equipotential lines, which are concentric ellipses in anisotropic traps (with different trapping frequencies in $x-$ and $y-$ axis, $\omega_x/\omega_y=1.25$). We find that the dynamics of vortices still depends on the configuration of the quadrupole structure in anisotropic systems. The parameter ranges can also be determined according to Fig. \ref{fig5}. However, $r$ which is chosen to identify the dynamical regimes is not the radial position of the vortices as in the isotropic systems, but the major radius or minor radius of the ellipse decided by the vortex positions, since the TF-radius of the condensate in each direction is different. By releasing the trapping potential in one direction (both directions) suddenly or adiabatically (the trapping frequencies are decreased linearly within $10t_0$) to a smaller value (form $2\pi\times 5$ \textrm{Hz} to $2\pi\times3$ \textrm{Hz} or $2\pi\times4$ \textrm{Hz} in our calculations), we confirm that the dynamics of quadrupole structures still obeys the same rule, which supports the conclusion in Ref. \cite{PRA.83.011603}, where they showed that the sudden reduction in the number of atoms following expansion is expected to excite a small-amplitude (a few percent) breathing mode, with only slight alteration of vortex trajectories, which means that the dynamics of vortices is robust under the expansion of condensates.

To examine the robustness of the quadrupole dynamics, we introduce quantum fluctuations into the systems we studied above as stated in Sec.2. We find that in the recombination regime with fluctuations the trajectories of vortices are nearly the same as those in the unperturbed systems during the first few periods. However, after that the quadrupole structure is eventually destroyed and the vortices take chaotic trajectories as shown in Fig. \ref{fig6}a. In the exchange regime with fluctuations, the dynamics of the vortices still obeys the rule of the unperturbed systems but with an additional rotation of the quadrupole structure as shown in Fig. \ref{fig6}b. This rotation also occurs if a dipole is placed asymmetrically about the origin \cite{CPAA.10.1589}. Moreover the vortices will not maintain a rectangular configuration anymore, leading to the destroy of the quadrupole structure and chaotic trajectories of vortices eventually. In the annihilation regime the quantum fluctuations introduced will not change the dynamics of the system since the vortices annihilate permanently as soon as they meet at the trap centre during the first period.

In the frame of zero temperature mean-field theory we expect that quantum fluctuations will not change the dynamics of the system dramatically especially for the short-time evolution. However, symmetries of the system are affected by quantum fluctuations, which leads to chaotic motion of vortices for the long-time evolution. As we focused on the first few periods of quarupole dynamics in our discussions, it shows that our results are robust, i.e. quantum fluctuations do not change the dynamical regimes of the system we studied at least for the short-time evolution. This also demonstrates that vortices have very stable topological structures.

We note that even though our calculations are limited at zero temperature, the vortices of the quadrupole cluster do deviate from their original locations during the periodical movement, which is similar to the situation in finite temperatures. The dynamics of the next period is fully decided by the fact that which region in Fig. \ref{fig5} the radial positions (major radius or minor radius) of the vortices will fall into. Transition from one regime to another of the system dynamics may occur during the time evolution if the vortices do not annihilate permanently, which means that the initial configurations whose dynamics is ruled by ``recombination'' in the first period may transit to the other two regimes in the following period. And this may finally result in expulsion of the vortices from the condensate.

\section*{Conclusion}\label{sec3}

In this paper, we have shown that there are three characteristic dynamical regimes in a trapped BEC according to the varying radial positions of vortices in a quadrupole structure. The closer the vortices are located to the periphery of the condensate, the higher the kinetic energy they will obtain when approaching the centre of the condensate, which will compete with the interatomic interactions and the vortex interactions and leading to distinct dynamical regimes. In the recombination regime the kinetic energy is too small to overcome the interatomic interactions, which prohibits the formation of soliton rings, and the vortices recombine into dipoles. In the annihilation regime the ring structure arises and expands fast out of the condensate before it decays into vortex dipoles, while in the exchange regime the emerged soliton ring can break into dipoles with charge flipping due to higher kinetic energy possessing by each vortex. These are also valid for anisotropic systems and systems with quantum fluctuations during the short-time evolution, where the symmetry of the system is altered. For the long-time evolution, the symmetry of the system is important for maintaining the rectangular configuration of a quadrupole structure.

By providing the relation between TF-radius of the condensate and the configuration of the vortex quadrupoles, our results disclose detailed dynamics of vortex quadrupole in a trapped BEC and a road map to distinguish them, and can be extended to other special structures, which helps to study more complex vortex settings. The numerical simulations also demand new theoretical models about the generation and expanding speed of soliton rings, by the combination of individual vortices, which are crucial to understand the dynamics of vortex clusters.


\section*{Acknowledgement}

This research was supported by National Natural Science Foundation of China under grant Nos.11347025, 11434015, 61227902, 61378017, NKBRSFC under grants Nos. 2012CB821305, SKLQOQOD under grants No. KF201403, SPRPCAS under grants No. XDB01020300 and Science Foundation of Northwest University under grant No. 13NW16.

\typeout{}
\begin{singlespace}
\bibliographystyle{unsrt}
\bibliography{ThesisEx}
\end{singlespace}

\end{document}